\documentclass{article}

\usepackage{spconf}
\usepackage{amsmath, graphicx, multirow}

\def\figurewidth{0.44}

\title{Future Vector Enhanced LSTM Language Model for LVCSR}
\name{Qi Liu, Yanmin Qian, Kai Yu}
\address{
 Key Lab. of Shanghai Education Commission for Intelligent Interaction and Cognitive Engineering \\
 SpeechLab, Department of Computer Science and Engineering \\
 Brain Science and  Technology Research Center \\
 Shanghai Jiao Tong University, Shanghai, China \\
Emails: \{liuq901, yanminqian, kai.yu\}@sjtu.edu.cn}

\begin{document}

\maketitle

\begin{abstract}
	Language models (LM) play an important role in large vocabulary continuous speech recognition (LVCSR). However, traditional language models only predict next single word with given history, while the consecutive predictions on a sequence of words are usually demanded and useful in LVCSR. The mismatch between the single word prediction modeling in trained and the long term sequence prediction in read demands may lead to the performance degradation. In this paper, a novel enhanced long short-term memory (LSTM) LM using the future vector is proposed. In addition to the given history, the rest of the sequence will be also embedded by future vectors. This future vector can be incorporated with the LSTM LM, so it has the ability to model much longer term sequence level information. Experiments show that, the proposed new LSTM LM gets a better result on BLEU scores for long term sequence prediction. For the speech recognition rescoring, although the proposed LSTM LM obtains very slight gains, the new model seems obtain the great complementary with the conventional LSTM LM. Rescoring using both the new and conventional LSTM LMs can achieve a very large improvement on the word error rate.
\end{abstract}
\noindent\textbf{Index Terms}: speech recognition, language model, recurrent neural network, n-best rescoring

\section{Introduction}
Language model plays an important role in LVCSR. N-gram \cite{Broder1997, Dunning1994} has been widely used in the LVCSR system for a long time. However, n-gram only uses limited histories which is hard to deal with long context sequences. RNN and LSTM language models \cite{Hochreiter1997, Mikolov2010} which can store the whole history of the sequence have been proposed to deal with this problem and obtained great success in many fields \cite{Chen2015_, Hu2014}.

However, many sequence level tasks including machine translation \cite{Brants2007}, speech recognition \cite{Chen2015} and handwriting recognition \cite{Liu2015} need long term sequence prediction, while the traditional RNN language model only predicts single word one by one. According to \cite{Ranzato2016}, there is a gap between the common used word level metric perplexity (PPL) for language model evaluation and the true sequence level metric such like BLEU score in machine translation \cite{Papineni2002} and word error rate (WER) in speech recognition \cite{Klakow2002}.

%Several researches \cite{He2016} \cite{Ranzato2016} \cite{Wiseman2016} \cite{Pichotta2016} have focused on this problem. These works use bidirectional LSTM, reinforcement learning, sequence training or encoder-decoder system to deal with this problem.

Several researches have been done to deal with this problem. \cite{He2016, Shi2013, Arisoy2015} researched on training bidirectional LSTM language model, which can retrieve the the information not only from the past context but also the future context. \cite{Ranzato2016} combined reinforcement learning and deep learning together, directly trained the neural network with the estimated BLEU score. \cite{Wiseman2016, Pichotta2016} applied sequence to sequence training method on language model.

% However, the above systems are not designed for long term sequence prediction. Long term sequence prediction need to generate the whole sequence with given history. Normal RNN or LSTM language model can generate the word one by one until the whole sequence is complete. However, due to it is lack of sequence level information, the performance may be degraded.

In this paper, an novel enhanced LSTM language model has been proposed. Enhanced LSTM language model predicts not only a single word, but also the whole future of the input sequence. It is believed that enhanced LSTM language model can perform well with more sequence level information.

% Enhanced LSTM language model uses a reversed LSTM language model to extract its activation values of the last hidden layer as bottleneck features \cite{Gehring2013} which can embed the future of a sequence. These activation values are called future vectors. These future vectors which contain sequence level information will be used to train the enhanced LSTM language model. The experiments show that the enhanced LSTM language model performs well on the sequence prediction task. It is also observed that in n-best rescoring task, the WER can get a very large improvement by the combination on the normal and enhanced LSTM language model.

Enhanced LSTM language model trains a reversed LSTM language model. And the activation values of the last hidden layer of this reversed LSTM are used as bottleneck features \cite{Gehring2013} which can embed the future of the sequence. These bottleneck features are called future vectors of the sequence.

These future vectors which contain sequence level information will be used to train the enhanced LSTM language model. The model will be trained by not only to predict the next word but also the future vector. The predicted future vector will also be the input feature to predict the next word.

The experiments show that the enhanced LSTM language model performs well on the sequence prediction task. It is also observed that in n-best rescoring task, the WER can get a very large improvement by the combination on the normal and enhanced LSTM language model.

The rest of the paper is organized as follows, section \ref{sec:bg} is the background. Section \ref{sec:sys} indicates the methodology of enhanced LSTM language model and section \ref{sec:exp} shows the experimental setup and results. Finally, conclusion will be given in section \ref{sec:conc} and discussion can be found in section \ref{sec:discu}.

\section{Background} \label{sec:bg}
\subsection{Long Short-Term Memory}

RNN \cite{Elman1990} is the neural network with cycles in its structure, which is effective in dealing with sequential data. Suppose there is a sequence of data $x_1, x_2, \dots, x_T$ as the input and let $h_1, h_2, \dots, h_T$ be the output of one RNN, the most commonly used RNN formula looks like $$h_t=f(W_xx_t+W_hh_{t-1}+b).$$ where $W_x$ and $W_h$ are weight matrix parameters, $b$ is the bias and $f$ is the activation.

\begin{figure}
	\centering
	\includegraphics[width=\figurewidth\textwidth]{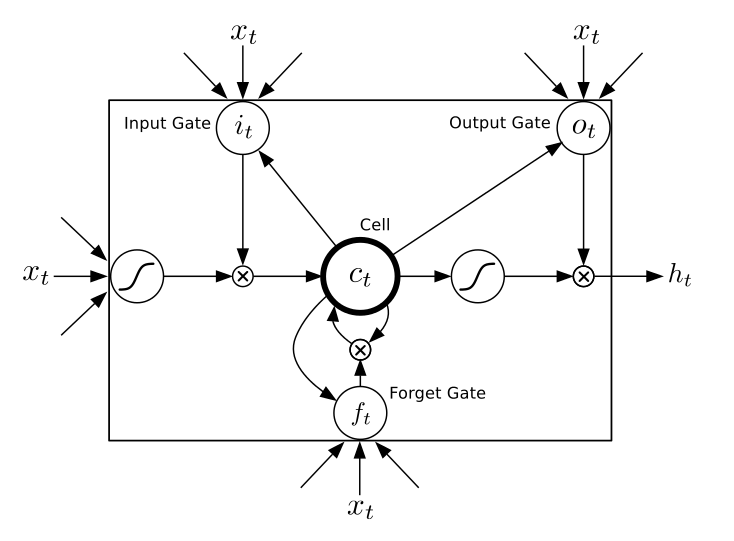}
	\caption{One LSTM memory cell \cite{Graves2014}. There are three gates (input gate, output gate and forget gate) in each cell to control the data flow. In practice, $h_{t-1}$ will also be the input to the cell together with $x_t$.}
	\label{fig:lstm}
\end{figure}

Due to gradient vanishing and explosion problems \cite{Hochreiter2001, Pascanu2013}, LSTM \cite{Hochreiter1997}, which is a unit structured RNN, has been used to replace the traditional RNN. LSTM-RNN shows better performance \cite{Gers2002, Graves2006, Chan2016}, and the LSTM formula is shown below:

\begin{align*}
{i_t}&={\sigma(W_{xi}x_t+W_{hi}h_{t-1}+W_{ci}c_{t-1}+b_i)} \\
{f_t}&={\sigma(W_{xf}x_t+W_{hf}h_{t-1}+W_{cf}c_{t-1}+b_f)} \\
{m_t}&={\tanh(W_{xc}x_t+W_{hc}h_{t-1}+b_c)} \\
{c_t}&={f_t\cdot c_{t-1}+i_t\cdot m_t} \\
{o_t}&={\sigma(W_{xo}x_t+W_{ho}h_{t-1}+W_{co}c_t+b_o)} \\
{h_t}&={o_t\cdot\tanh(c_t)}.
\end{align*}
where $W_{**}$ are the weight matrix parameters, $b_*$ are the bias and $\sigma$ is the sigmoid function. The detail of its structure can be found in Figure \ref{fig:lstm}.

\subsection{LSTM Language Model} \label{sec:lstm_lm}

\begin{figure}
	\centering
	\includegraphics[width=\figurewidth\textwidth]{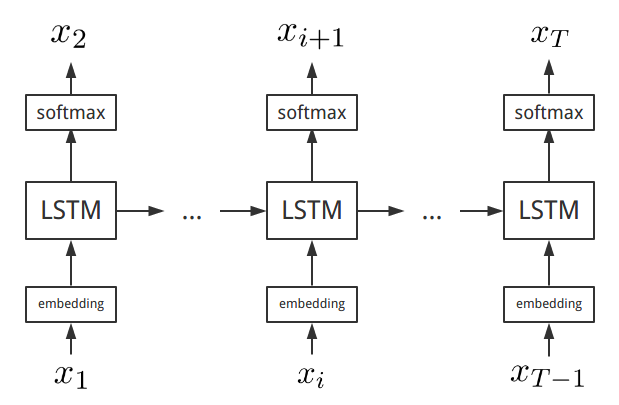}
	\caption{The structure of LSTM language model. Here $x_1, x_2, \ldots, x_T$ is the input sequence.}
	\label{fig:lstm_lm}
\end{figure}

LSTM language model uses the current word as the input and the next word as the output. In detail, suppose $x_1, x_2, \dots, x_T$ is the input sequence, $x_i$ is the $i$-th word, and the vocabulary size is $n$. The input layer of the LSTM is a word embedding layer with size $n$, and the output layer of the LSTM is a softmax layer with size $n$. The detail formula is shown below:
\begin{align*}
\bar{x_i}&=f(x_i) \\
h_i&=\text{LSTM}(\bar{x_i}, h_{i-1}) \\
p_i&=\text{softmax}(Wh_i+b) \\
x_{i+1}&=\arg\max{p_i},
\end{align*} 
where $f$ represents the word embedding and $W, b$ are the network parameters. Figure \ref{fig:lstm_lm} shows the structure of LSTM language model. At the $i$-th time step, $x_i$ is the input to the LSTM, and the output value $p_i=(p_i^{(1)}, p_i^{(2)}, \ldots, p_i^{(n)})$ is considered to be the probability of observe each word at time step $i+1$, i.e. $$p(x_{i+1}|x_1, x_2, \ldots, x_i)=p_i^{(x_{i+1})}.$$ 

To train the LSTM language model, the cross entropy (CE) of output distribution $p_i$ and the ground truth distribution $$g_i=(0, \ldots, 0, 1, 0, \ldots, 0|1 \text{ at position } x_{i+1})$$ will be used as the criterion to train the network, i.e. the loss function is $$\mathcal{L}=\text{CE}(g_i,p_i)=-\sum_{j=1}^ng_i^{(j)}\log p_i^{(j)}.$$

\section{Methodology} \label{sec:sys}

\begin{figure}
	\centering
	\includegraphics[width=\figurewidth\textwidth]{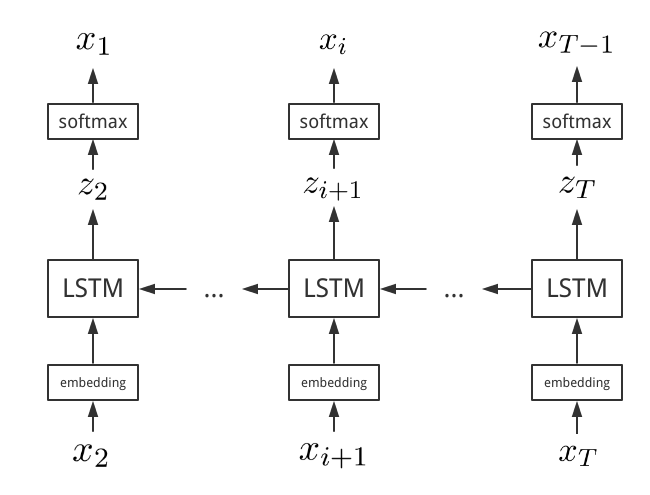}
	\caption{The structure of future vector extractor. Here $x_1, x_2, \ldots, x_T$ is the input sequence and $z_2, z_3, \ldots, z_T$ are the extracted future vectors.}
	\label{fig:lstm_rev}
\end{figure}

\subsection{Future Vector Extraction} \label{sec:fv_ext}
Traditional LSTM language models only predict a single word for the given history, which may lose information about the whole future. In contrast the rest of the sequence will be embedded into a sequence vector in the new proposed enhanced LSTM language model. This sequence vector, which is called future vector in this paper, contains the information about all the sequence future.

There are several ways \cite{Le2014, Iyyer2015, Kalchbrenner2014} to extract future vectors. What is needed here is that for a given input sequence, each suffix needs be embedded and the relationship among them must be kept. Therefore the method similar to \cite{Palangi2016} has been chosen. A normal LSTM language model with reversed input sequence order has been trained, which means this LSTM language model predicts the previous word with the given future. The future vector is extracted from the activation values of the last hidden layer in this reversed LSTM language model. Figure \ref{fig:lstm_rev} shows the detailed structure and the formula is shown below.
\begin{align*}
\bar{x_i}&=f(x_i) \\
z_i&=\text{LSTM}(\bar{x_i}, z_{i+1}) \\
p_i&=\text{softmax}(Wz_i+b) \\
x_{i-1}&=\arg\max p_i,
\end{align*}
where $f$ is the word embedding and $W, b$ are model parameters. $z_2, z_3, \ldots, z_T$ are the extracted future vectors.

\begin{figure}
	\centering
	\includegraphics[width=\figurewidth\textwidth]{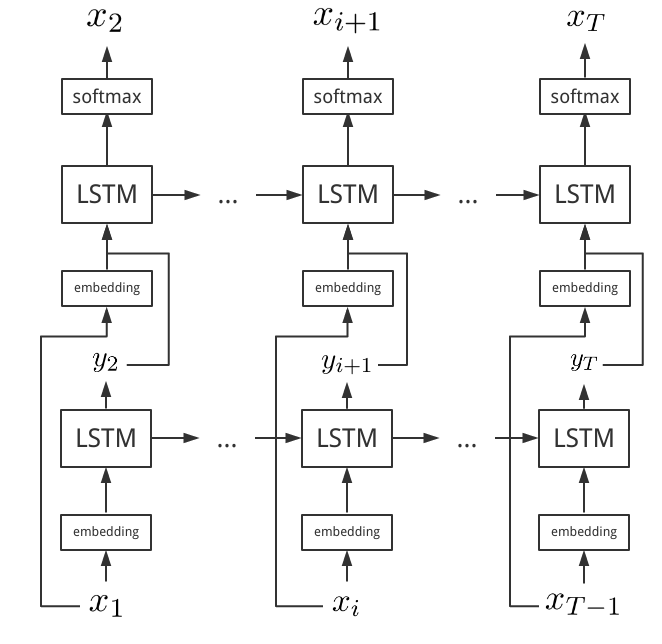}
	\caption{The structure of enhanced LSTM language model. Here $x_1, x_2, \ldots, x_T$ is the input sequence and $y_2, y_3, \ldots, y_T$ are the predicted future vectors. In practice, the two LSTM networks are trained separately.}
	\label{fig:lstm_sys}
\end{figure}

\subsection{Enhanced LSTM Language Model} \label{sec:en_lstm_lm}
Future vectors cannot be directly used to train a language model. For a input sequence $x_1, x_2, \ldots, x_T$ and its future vectors $z_1, z_2, \ldots, z_T$, only history $x_1, x_2, \ldots, x_i$ are known while the language model is trying to predict word $x_{i+1}$. However, the future vector $z_{i+1}$ is a function of unknown future $x_{i+1}, x_{i+2}, \ldots, x_T$ which is impossible to be generated.

One additional LSTM network has been trained to solve this problem. This network is similar to normal LSTM language model but predicts the future vector rather than the next word. The detailed formula is
\begin{align*}
\bar{x_i}&=f(x_i) \\
h_i&=\text{LSTM}(\bar{x_i}, h_{i-1}) \\
y_{i+1}&=Wh_i+b
\end{align*}
where $f$ is word embedding and $W, b$ are network parameters.
The criterion to train this network is the mean squared error (MSE) between the future vector prediction $y_i$ and the truly extracted future vector $z_i$ described in section \ref{sec:fv_ext}, i.e. the error function is $$\mathcal{L}=\text{MSE}(y_i,z_i)=\frac{1}{m}\sum_{j=1}^m(y_i^{(j)}-z_i^{(j)})^2,$$ where $m$ is the dimension of future vector.

$y_i$ is a function of $x_1, x_2, \ldots, x_{i-1}$ which means it can be directly used to train a language model. In enhanced LSTM language model, $y_{i+1}$ will be combined together with $x_i$ as the new input of the LSTM language model, i.e.
\begin{align*}
\bar{x_i}&=f(x_i) \\
h_i&=\text{LSTM}(\bar{x_i}, y_{i+1}, h_{i-1}) \\
p_i&=\text{softmax}(Wh_i+b) \\
x_{i+1}&=\arg\max{p_i},
\end{align*} 
where $f$ indicates the word embedding and $W, b$ are network parameters. The criterion is CE which is the same as normal LSTM language model in section \ref{sec:lstm_lm}.
The details structure is illustrated in figure \ref{fig:lstm_sys}.

Enhanced LSTM language model has more input, the future vector $y_i$, to predict the next word compared with the normal LSTM language model. This results an enhanced LSTM language model which has the power ability to modeling future sequence level information.

\begin{figure}
	\centering
	\includegraphics[width=\figurewidth\textwidth]{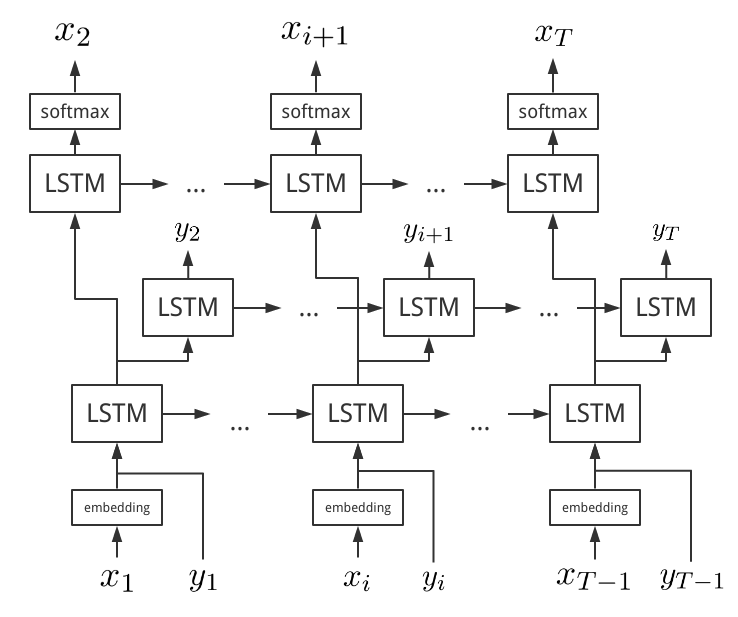}
	\caption{The structure of multi-task enhanced LSTM language model. Here $x_1, x_2, \ldots, x_T$ is the input sequence and $y_2, y_3, \ldots, y_T$ are the predicted future vectors. $y_1$ is a zero vector. In practice, the three LSTM networks are trained together.}
	\label{fig:lstm_mt}
\end{figure}

\subsection{Multi-task Enhanced LSTM}
Enhanced LSTM language model has two networks, one is future vector prediction LSTM and the other one is language model LSTM. It is observed that these two networks can be trained together. Multi-task training \cite{Yin2016, Xue2007, Reichart2008} is a suitable method for joint training.

The prediction of next word and corresponding future vector can be optimized at the same time in the multi-task enhanced LSTM language model. The predicted future vector will also be the input like the non multi-task version. The detailed formula is here,
\begin{align*}
\bar{x_i}&=f(x_i) \\
h_i&=\text{LSTM}(\bar{x_i}, y_{i}, h_{i-1}) \\
u_i&=\text{LSTM}(h_i, u_{i-1}) \\
y_{i+1}&=W_uu_i+b_u \\
v_i&=\text{LSTM}(h_i, v_{i-1}) \\
p_i&=\text{softmax}(W_vv_i+b_v) \\
x_{i+1}&=\arg\max{p_i},
\end{align*}
where $f$ is the word embedding and $W_*, b_*$ are network parameters. The two criteria to train this multi-task network is MSE for future vector prediction and CE for word prediction which also have been used for non multi-task version in section \ref{sec:en_lstm_lm}, i.e. the loss function is $$\mathcal{L}=\text{CE}(g_i,p_i)+\lambda\text{MSE}(y_{i+1},z_{i+1}),$$ $\lambda=1.0$ in this implementation. The structure is Figure \ref{fig:lstm_mt}.

Multi-task enhanced LSTM language model can get not only explicit sequence level information from the input but also the implicit sequence level information from the future vector prediction.

\begin{table}[bht]
	\begin{center}
		\begin{tabular}{|c|c|c|}
			\hline 
			\bf Model & \bf Input & \bf Output \\ 
			\hline
			LSTM & $x_i$ & $x_{i+1}$ \\
			\hline
			\multirow{2}{*}{FV} & $x_i$ & $y_{i+1}$ \\
			& $x_i, y_{i+1}$ & $x_{i+1}$ \\
			\hline
			MT-FV & $x_i, y_i$ & $x_{i+1},y_{i+1}$ \\
			\hline
		\end{tabular}
	\end{center}
	\caption{Brief comparison among three LSTM language model structures. FV indicates the future vector enhanced LSTM, and MT-FV indicates the future vector enhanced LSTM with multi-task training. $x_*$ indicates the original input sequence and $y_*$ is the predicted future vector.}
	\label{tab:sys}
\end{table}

In table \ref{tab:sys}, a briefly comparison of structures among normal LSTM, enhanced LSTM and multi-task enhanced LSTM language model has been shown.

% \begin{table}
% 	\begin{center}
% 		\begin{tabular}{|c|c|c|}
% 			\hline 
% 			\bf Corpus & \bf \# of Sentence & \bf Vocabulary Size \\ 
% 			\hline
% 			PTB & 49199 & 10000 \\
% 			\hline
% 			SMS & 403218 & 40697 \\
% 			\hline
% 			%% FSH & 2481808 & 31001 \\
% 			%\hline
% 		\end{tabular}
% 	\end{center}
% 	\caption{Statistics of two corpora. SMS indicates short message corpus.}
% 	\label{tab:data}
% \end{table}

\begin{table*}
	\centering
	\begin{tabular}{|c|c|c|c|c|c|c|c|}
		\hline
		\multirow{2}{*}{\bf Corpus} & \multirow{2}{*}{\bf Model} & \multirow{2}{*}{\bf Perplexity} &
		\multicolumn{5}{|c|}{\bf BLEU Score} \\
		\cline{4-8}
		& & & \bf 0 & \bf 1 & \bf 2 & \bf 3 & \bf 5 \\
		\hline
		\multirow{3}{*}{PTB} & LSTM & 122 & 0.076 & 0.083 & 0.092 & 0.097 & 0.106 \\
		& FV-LSTM & 120 & 0.081 & 0.094&0.099&0.104&0.112 \\
		&MT-FV-LSTM & 120 &0.076&0.084&0.091&0.098&0.105 \\
		\hline
		\multirow{3}{*}{SMS} & LSTM & 105&0.179&0.222&0.241&0.262&0.277 \\
		& FV-LSTM & 102&0.212&0.243&0.261&0.273&0.285 \\
		&MT-FV-LSTM & 104&0.187&0.225&0.243&0.265&0.284 \\
		\hline
		%\multirow{3}{*}{FSH} & LSTM & 62.5 & 0.139&0.137&0.140&0.147&0.154 \\
		%\cline{2-8}
		% &MT-EN-LSTM & 63.0 &0.139&0.135&0.138&0.147&0.156 \\
		%\cline{2-8}
		% & EN-LSTM & 63.5 & 0.140 & 0.137&0.144&0.149&0.157 \\
		%\hline				
	\end{tabular}
	\caption{PPL and BLEU comparison of sequence prediction task. FV-LSTM indicates the future vector enhanced LSTM, and MT-FV-LSTM indicates the future vector enhanced LSTM with multi-task training. The number below BLEU score is the length of history.
	}
	\label{tab:bleu}
\end{table*}

\section{Experiments} \label{sec:exp}
\subsection{Experimental Setup}
The experiments are designed to evaluate the performance of the proposed enhanced LSTM language model. The experiments uses two corpora including PTB English corpus and short messages Chinese corpus. PTB corpus contains 49199 utterances and Chinese short messages corpus has 403218 utterances. The vocabulary size is 10000 and 40697 respectively. The experiments used almost the same structure in all the systems. All the LSTM block in Figure \ref{fig:lstm_lm}, \ref{fig:lstm_rev} and \ref{fig:lstm_sys} is a stacked three hidden layers LSTM. In Figure \ref{fig:lstm_mt}, the multi-task network has two hidden LSTM layers in shared part and one hidden LSTM layer in separate part. All the LSTM hidden layers contains 300 cells.

Both sequence prediction and speech recognition n-best rescoring will be evaluated, and the BLEU score and WER are used respectively.

\subsection{Experimental Results of Sequence Prediction}
The results of sequence prediction can be found in Table \ref{tab:bleu}. For each test sequence, five different lengths (0, 1, 2, 3 and 5) of history were used. The BLEU score which is calculated between the ground truth and prediction is used as the evaluation metric.

It can be observed that the PPL keeps almost the same in all the three systems. It is not surprising due to the enhanced LSTM language model is focused on the improvement of sequence level performance but PPL is a word level metric. However, the enhanced LSTM language model performs consistent better on BLEU score with different history lengths. These demonstrate that the enhanced LSTM language model can retrieve more sequence level information and get better result on sequence level metric.

% It is not surprised that the PPL of the three models are almost the same on all the corpora. Enhanced LSTM language model focuses on improvement of sequence level performance. PPL, as a word level metric, remains almost unchanged indicates that enhanced LSTM language model does not reduce the word level performance while trying to add more sequence level information.

% The BLEU score shows that enhanced LSTM language model truly performs better on sequence level metric. Compared with normal LSTM language model, the enhanced LSTM language model gets a better BLEU score with all the lengths of given history. Multi-task LSTM language model gets almost the same performance on PTB corpus, but performs well on short message corpus with little given history. It also can be observed that enhanced LSTM language model gets more improvements when the length of given history is short. It is believed that the reason is long history can give enough implicit sequence level information for prediction of normal LSTM language model. However, enhanced LSTM language model can retrieve explicit sequence level information even when the history is short, which gets a much higher BLEU score than normal LSTM language model when the history is short.
% For Fisher-Switchboard corpus, the task is more difficult therefore all the methods get not very good results when the history is short. However, when the history is enough, enhanced LSTM language model performs slightly better than normal LSTM language model.

To give a better understanding on the results comparison, an example has been given with the history "Japan however has", and the results of three models (traditional LSTM, enhanced LSTM, multi-task enhanced LSTM) are shown as below:
\begin{itemize}\setlength{\itemsep}{0pt}
	\item Japan however has a N of its million;
	\item Japan however has been a major brand for the market;
	\item Japan however has been a major part of the company.
\end{itemize}
It can be observed that the enhanced LSTM language model gives more natural results on sequence prediction.

\subsection{Experimental Results of N-best Rescoring}
The Chinese SMS corpus is used to do speech recognition n-best rescoring. In the speech decoding stage for each audio, the sequences with the 100 highest probability will be generated. In the language model rescoring the language model score will be re-calculated by LSTM and enhanced LSTM language models, and the best path is obtained by combining both the language model score and acoustic model score. The WER comparison of n-best rescoring with different LSTM language models is given in Table \ref{tab:fsh}.

\begin{table}[bht]
	\begin{center}
		\begin{tabular}{|c|c|}
			\hline 
			\bf Model & \bf WER \\
			\hline
			3-gram & 12.85 \\
			\hline
			LSTM & 11.39 \\
		%	\hline
			FV & 11.35 \\
		%	\hline
            FV-MT & 11.29 \\
			\hline
            LSTM + FV & 10.84 \\
        %    \hline
            LSTM + FV-MT & 10.75 \\
            \hline
            LSTM + FV + FV-MT & 10.65 \\
            \hline
		\end{tabular}
	\end{center}
	\caption{WER (\%) comparison of speech recognition n-best rescoring on Chinese SMS corpus. FV indicates the future vector enhanced LSTM, and MT-FV indicates the future vector enhanced LSTM with multi-task training. All the models use equally interpolated weights.}
	\label{tab:fsh}
\end{table}

It can be observed that all LSTM language models can get a large improvement over the 3-gram language model, and the new proposed LSTM language model enhanced with future vector only get a slight gain compared to the traditional LSTM language model in the single model rescoring. However, when implementing the multiple LSTM language models rescoring shown as the bottom part of Table \ref{tab:fsh}, the new proposed future vector enhanced LSTM language models seem to own the huge complementary with the traditional LSTM language model. Rescoring using both the new and conventional LSTM language model together can achieve another significant improvement compared to the single LSTM language model rescoring.

\section{Conclusion} \label{sec:conc}
Traditional LSTM language model only predicts a single word with the given history. However, LVCSR need sequence level predictions. This mismatch may cause the degradation on the performance. In this paper, a novel enhanced LSTM language model has been proposed. Enhanced LSTM language model retrieves sequence level information from future vector which is a special kind of sequence vector. Therefore enhanced LSTM language model is able to predict long term future rather than immediate word. The experiments demonstrated that the proposed enhanced LSTM language model with future vector performs well on n-best rescoring than the traditional LSTM language model, and there is a huge complementary within the new and normal LSTM language models. The results of sequence prediction also indicate that the enhanced LSTM language model can be used on other sequence level tasks.

\section{Discussion} \label{sec:discu}
Enhanced LSTM language model is an enhanced version of traditional LSTM language model, it is still a word level supervised neural network model. This is an advantage that in the pipeline of other applications, traditional LSTM language model can be straightforward replaced by enhanced LSTM language model. However, this makes the performance of enhanced LSTM language model relies on the information contains in the future vector and prediction accuracy of future vector prediction network. If the extracted future vector or predicted future vector are not generated properly, the enhanced LSTM language model system may give worse results than normal LSTM language model. Thus, the future work is listed here,
\begin{enumerate}\setlength{\itemsep}{0pt}
	\item add gate to the network to control the scale of word level and sequence level information;
	\item try other ways to extract future vector;
	\item implement different methods to predict future vector;
	\item use reinforcement learning to train the network directly with the sequence level evaluation metric;
	\item use other sequence level tasks to test enhanced LSTM language model.
\end{enumerate}

\section{Acknowledgement}
This work was supported by the Shanghai Sailing Program No. 16YF1405300, the China NSFC projects (No. 61573241 and No. 61603252) and the Interdisciplinary Program (14JCZ03) of Shanghai Jiao Tong University in China. Experiments have been carried out on the PI supercomputer at Shanghai Jiao Tong University.

\bibliographystyle{IEEEbib}

\bibliography{paper}

\end{document}